\documentclass[prl, showpacs, twocolumn, floatfix]{revtex4}

\usepackage{graphicx}
\usepackage{amsmath, amsfonts, amssymb, bm}
\usepackage[]{psfrag}

\psfragscanoff
\def\omfeld{\bar\omega}
\def\omvak{\omega_k}

\begin{document}
%%%%%%%%%%%%%%%%%%%%%%%%%%%%%%%%%%%%%%%%%%%%%%%%%%%%%
%%%%%%%%%%%%%%%%%%%%%%%%%%%%%%%%%%%%%%%%%%%%%%%%%%%%%
%\title{Suppressing spontaneous emission via an external low-frequency field}
\title{Spontaneous emission suppression on arbitrary atomic transitions}
\author{J\"org \surname{Evers}}
\email{evers@physik.uni-freiburg.de}
\author{Christoph H. \surname{Keitel}}
\email{keitel@physik.uni-freiburg.de}
\affiliation{Theoretische Quantendynamik, Fakult\"at f\"ur Physik, Universit\"at Freiburg, Hermann-Herder-Stra{\ss}e 3, D-79104 Freiburg, Germany}
\date{\today}

%%%%%%%%%%%%%%%%%%%%%%%%%%%%%%%%%%%%%%%%%%%%%%%%%%%%%
%%%%%%%%%%%%%%%%%%%%%%%%%%%%%%%%%%%%%%%%%%%%%%%%%%%%%
\begin{abstract}
We propose a very simple scheme to slow down the usual exponential decay of upper state population in an atomic two level system considerably. The scheme makes use of an additional intense field with frequency lower than the total decay width of the atomic transition. This allows for additional decay channels with the exchange of one or more low-frequency photons during an atomic transition. The various channels may then interfere with each other. The intensity and the frequency of the low-frequency field  are shown to act as two different control parameters modifying the duration and the amount of the population trapping. An extension of the scheme to include transitions to more than one lower state is straightforward.
\end{abstract}

\pacs{42.50.Lc, 42.50.Ct, 42.50.Hz}

\maketitle
%%%%%%%%%%%%

Spontaneous decay is one of the main limiting factors for a possible application of many physical processes. Examples are high frequency lasers, the storage and processing of quantum information, or secure information transmission using quantum effects. In all of these examples, spontaneous decay or the decoherence caused by it limits the experimental feasibility. Thus it is one of the main goals of quantum optics to find possibilities to strongly reduce or even inhibit spontaneous decay. Various schemes have been proposed to overcome the spontaneous decay. 

One possibility is to make use of the quantum Zeno effect \cite{zeno}. This is based on the measurement postulate which states that a system is projected 
into one of its eigenstates upon a measurement. If the measurements are repeated rapidly, the system evolution may effectively be stopped. The
technical conditions on the brevity of the pulses though may not always be easily fulfilled for every transition.
Another method is the modification of the vacuum surrounding the given atomic system, e.g. by an optical cavity \cite{cavity}. 
By this the mode density at the frequency of the atomic transition may be altered such that the given transition is suppressed.
Here the control of the environmental modes with cavities is rather challenging in reality.
Thirdly quantum interference effects may be used to find superpositions of more than one upper state which are stable \cite{interference1,shapiro} 
or almost stable \cite{interference2} against spontaneous decay. In spite of the conceptual beauty, the disadvantage here often is the difficulty to 
provide convenient atomic systems which fulfill all conditions for interference to be present.

In this letter, we propose a very simple and practical scheme to slow down the usual exponential decay of the upper state population of any atomic transition
considerably.  An intense low-frequency field is applied to the system such that the frequency of the low-frequency field is lower than the decay width 
of the atomic system. The intensity is assumed to be constant for the duration during which the decay shall be controlled. Then in addition to the usual 
decay to the ground state, the atom may interact with the low-frequency field and exchange one or 
more low-frequency photons during an atomic transition. This effectively creates different decay paths from the upper to the lower atomic state which 
may be interpreted as an upper state multiplet decaying to a common ground state. Neighboring states in the multiplet are separated by the low-frequency 
field frequency. As this frequency is low as compared to the decay width of the atom, the upper states of this multiplet are closely 
spaced. Thus quantum interference amongst them is possible, which is shown to account for the modification of the usual exponential decay.
The analysis is carried out for a two-level system but the proposal may be implemented equally simple for systems involving multiple transitions as long
as the frequency of the applied field is smaller than each of the decay rates.

%%%%%%%%
\begin{figure}[b]
\psfrag{a}{$|e\rangle$}
\psfrag{b}{$|g\rangle$}
\psfrag{w1}{$\gamma^{(0)}$}
\psfrag{w2}{$\gamma^{(-1)}$}
\psfrag{w3}{$\gamma^{(+1)}$}
\includegraphics[height=3.5cm]{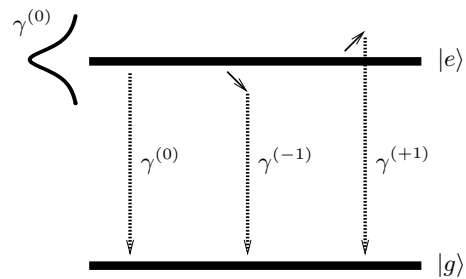}
\caption{\label{pic-system} Three of the possible decay channels on an arbitrary atomic transition exposed to an 
intense coherent field with a frequency not exceeding the spontaneous emission rate $\gamma^{(0)}$.
Destructive interference is shown to virtually inhibit the decay process.
}
\end{figure}
%%%%%%%%

The model system is shown in Fig. \ref{pic-system}. The atomic system consists of the states $\{ |e\rangle, |g\rangle \}$. The dotted lines denote spontaneous decay while the solid lines denote absorptions or emissions of low-frequency photons. The figure displays three of the possible transitions: the direct decay and the decay with annihilation/absorption of one low-frequency photon.

The model Hamiltonian can be written as the sum of the free Hamiltonian $H_0$ and the interaction part $H_I$
\begin{eqnarray*}
H_0 &=& \hbar \omega_g|g\rangle \langle g| +  \hbar \omega_e|e\rangle \langle e|
+ \hbar \omfeld b^\dagger b + \sum_k \hbar \omvak a_k^\dagger a_k \\
H_I &=& \hbar \sum_k \left ( g_k^* a_k \sigma_+ + g_k a_k^\dagger \sigma_-\right ) \times \\
&& \times (1 + \bar{g}_1 b + \bar{g}^*_1 b^\dagger + \bar{g}_2 (b)^2 + \bar{g}^*_2 (b^\dagger)^2 + \dots)
\end{eqnarray*}
where $\hbar \omega_g$ and $\hbar \omega_e$ are the respective energies of the atomic states, $b$ ($b^\dagger$) is the low-frequency photon annihilation (creation) operator, $a_k$ ($a^\dagger _k$) are the vacuum photon annihilation (creation) operators, $\sigma_+$ ($\sigma_-$) is the atomic raising (lowering) operator, $g_k$ are coupling constants to the various vacuum modes and $\bar{g}_i$ ($i=1,2,\dots$) are coupling constants representing the probability for a $i$-photon transition of the low-frequency field. $k$ is a multi index over all possible vacuum modes and polarizations. The $\bar{g}_i$ in general cannot be chosen independently; their relations depend on the specific atomic system used. However the overall probability of any interaction with the low-frequency field during an atomic transition, i.e. the relation of $\bar{g}_1$ to $1$ depends on the intensity of the low-frequency field and may be externally controlled.

Using the completeness relation $1 = \sum_{n=0}^{\infty} |n\rangle \langle n|$ where $|n\rangle$ is a low-frequency field state with $n$ photons and after shifting the summation indices $n$ the interaction Hamiltonian can be written as 
\begin{eqnarray*}
H_I =&\hbar& \sum_k  \sum_{n=0}^{\infty} \{ g_k^* \; a_k \; |e,n\rangle \langle g,n|  + g_k \;a_k^\dagger \;|g,n\rangle \langle e,n| + \\
&& g_k^* \;\bar{g}_1\; \sqrt{n}\; a_k \; |e,n-1\rangle \langle g,n| +\\
&& g_k \;\bar{g}_1^* \; \sqrt{n}\;a_k^\dagger \;|g,n\rangle \langle e,n-1| + \\
&& g_k^* \;\bar{g}_1^*\; \sqrt{n+1}\; a_k \; |e,n+1\rangle \langle g,n|  +\\
&& g_k \;\bar{g}_1 \; \sqrt{n+1}\;a_k^\dagger \;|g,n\rangle \langle e,n+1| + \\
&& g_k^* \;\bar{g}_2\; \sqrt{n(n-1)}\; a_k \; |e,n-2\rangle \langle g,n|  + \\
&& g_k \;\bar{g}_2^* \; \sqrt{n(n-1)}\;a_k^\dagger \;|g,n\rangle \langle e,n-2| + \\
&& g_k^* \;\bar{g}_2^*\; \sqrt{(n+1)(n+2)}\; a_k \; |e,n+2\rangle \langle g,n|  + \\
&& g_k \;\bar{g}_2 \; \sqrt{(n+1)(n+2)}\;a_k^\dagger \;|g,n\rangle \langle e,n+2| + \\
& & \dots \}
\end{eqnarray*}
Thus a two-level system with a low-frequency field which allows up to $N$ low-frequency photons to be emitted or absorbed during a spontaneous decay process is equivalent to an atomic system with one lower state $|b,n\rangle$ and $(2N+1)$ upper states $\{|e, n-N\rangle, \dots, |e, n+N\rangle \}$ without any additional fields but the vacuum. The value for $N$ depends on the actual atomic system used, the intensity of the low-frequency field and the relation of $\bar{\omega}$ to the atomic decay width. As will be shown later, $N=1$ already suffices for a considerable slow down of the population decay. However in general the effects become more pronounced with increasing $N$.

For a strong low-frequency field ($n\gg 1$) we can assume $n\approx n+1 \approx n-1$. Defining
\begin{eqnarray*}
&&g_k^{(0)} = g_k,  \qquad g_k^{(-1)} = g_k\; \bar{g}_1^*\; \sqrt{n}, \\
&&g_k^{(+1)} = g_k\; \bar{g}_1\; \sqrt{n+1}, \qquad g_k^{(-2)} = g_k \; \bar{g}_2^*\; \sqrt{n(n-1)} ,\\
&&g_k^{(+2)} = g_k \; \bar{g}_2\;\sqrt{(n+1)(n+2)}
\end{eqnarray*}
and equivalent for higher indices as coupling constants and 
\begin{eqnarray*}
\sigma^{(j)}_+ = \sum |e,n+j\rangle \langle g, n|, \qquad \sigma^{(j)}_- = (\sigma^{(j)}_+)^\dagger
\end{eqnarray*}
as generalized ladder operators, the interaction Hamiltonian summed over all low-frequency photon numbers in the interaction picture with respect to $H_0$ can be written as
\begin{eqnarray*}
V=&\hbar& \sum_k \{ (g_k^{(0)*} \; a_k \; \sigma_+^{(0)} e^{i\delta_k t}+ h.a. ) + \\
&& (g_k^{(-1)*} \; a_k \; \sigma_+^{(-1)} e^{i(\delta_k -\omfeld) t} + h.a. ) + \\
&& (g_k^{(+1)*} \; a_k \; \sigma_+^{(+1)} e^{i(\delta_k +\omfeld) t}+ h.a. ) + \\
&& (g_k^{(-2)*} \; a_k \; \sigma_+^{(-2)}e^{i(\delta_k -2\omfeld) t}  + h.a. ) + \\
&& (g_k^{(+2)*} \; a_k \; \sigma_+^{(+2)}e^{i(\delta_k +2\omfeld) t} + h.a. ) + \\
&& \dots \}
\end{eqnarray*}
with $\delta_k = \omega_e - \omega_g - \omega_k$ as detuning of the $k$th vacuum mode from the atomic transition frequency. Following the summation over all low-frequency photon numbers, the atomic ladder operator to and from one of the upper states $\sigma^{(j)}_\pm$ and the coupling constants $g_k^{(j)}$ ($j\in \{-N,\dots,N\}$)  are replaced by the corresponding semiclassical entities, for simplicity, without change of notation. 
%%%%%%%%
The system state vector can be written as 
\begin{equation}
|\Psi(t)\rangle = \left ( \sum _{j=-N}^{N} E^{(j)}(t) \; \sigma^{(j)}_+ \; |g\rangle |0\rangle \right )   + \sum _k G_k(t) \; a^\dagger _k \; |g\rangle |0\rangle \label{state}
\end{equation}
with time dependent coefficients $E^{(j)}(t)$ and $G_k(t)$ and
where $|0\rangle$ denotes the vacuum state, i.e. with no photons in the vacuum modes.
The equations of motion for the different coefficients of the atomic states can easily be obtained as in \cite{zhu} 
using the Schr\"odinger equation 
\begin{equation}
i\hbar \frac{d}{dt} |\Psi (t)\rangle = V |\Psi (t)\rangle . \label{schroedinger}
\end{equation}
Substituting Eq. (\ref{state}) in Eq. (\ref{schroedinger}), formally integrating the equation for $G_k(t)$ and 
resubstituting the result into the other equations, we obtain as equations of motion for the upper state 
coefficients after a Wigner-Weisskopf - like approximation \cite{oc}:
\begin{eqnarray}
\frac{d}{dt} E^{(j)} = - \frac{\gamma^{(j)}}{2} E^{(j)} - \sum_{\substack{l=-N\\l\neq j}}^{N} \frac{\gamma^{(l,j)}}{2}E^{(l)} e^{i(j-l)\bar{\omega}t}
\end{eqnarray}
with $\gamma^{(l,j)}=\sqrt{\gamma^{(l)}\gamma^{(j)}}$ for $j \in \{-N,\dots , N\} $. $\gamma^{(l)} = 2\pi |g^{(l)}|^2 D(\omega_e-\omega_g+l\bar{\omega})$ is the decay rate from the upper state $l$ to the lower state with mode density  $D(\omega)$ of the vacuum field.
With our assumption $\omfeld \ll (\omega_e - \omega_g)$ and because the upper states have parallel dipole moments, the system may give rise to quantum interference effects in the spontaneous decay. These are the reasons for the terms containing the $\gamma^{(l,j)}$.

For example, in the case $N=1$ the set of equations for the three upper states is 
\begin{eqnarray*}
\frac{d}{dt} E^{(-1)}(t) &=& - \frac{\gamma^{(-1)}}{2} E^{(-1)}(t) - \frac{\gamma^{(0,-1)}}{2}E^{(0)}(t) e^{-i\bar{\omega}t}  \\ 
&& - \frac{\gamma^{(+1,-1)}}{2}E^{(+1)}(t) e^{-2i\bar{\omega}t},  \\
\frac{d}{dt} E^{(0)}(t) &=& - \frac{\gamma^{(0)}}{2} E^{(0)}(t) - \frac{\gamma^{(-1,0)}}{2}E^{(-1)}(t) e^{i\bar{\omega}t}  \\
&& -\frac{\gamma^{(+1,0)}}{2}E^{(+1)}(t) e^{-i\bar{\omega}t}, \\
\frac{d}{dt} E^{(+1)}(t) &=& - \frac{\gamma^{(+1)}}{2} E^{(+1)}(t) - \frac{\gamma^{(-1,+1)}}{2}E^{(-1)}(t) e^{2i\bar{\omega}t}  \\ 
&&- \frac{\gamma^{(0,+1)}}{2}E^{(0)}(t) e^{i\bar{\omega}t}.  
\end{eqnarray*}
These equations can easily be solved e.g. using the computer algebra system {\it Mathematica}. The total upper population is then given by
\[
\Pi(t) = \sum _{j=-N}^{N} |E^{(j)}(t)|^2.
\]
%%%%%%%
\begin{figure}[t]
\psfrag{pop}[Bl]{$\Pi(t)$}
\psfrag{t}[c]{$t$ [units of $1/\gamma^{(0)}$]}
\includegraphics[width=8cm]{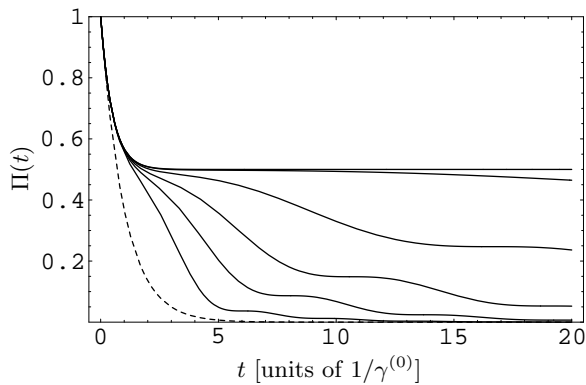}
\caption{\label{pop-omega}Upper state population for different values of $\bar{\omega}$. Starting from the top curve the chosen values are $\bar{\omega} = 0; 0.1; 0.3; 0.5; 0.7; 1$. The dashed curve is the usual exponential decay without the low-frequency field $\exp (-\gamma^{(0)} t)$. The other parameters are $N=1, \gamma^{(0)} = 1, \gamma^{(\pm 1)}=0.5$ and $E^{(0)}(0)=1$.}
\end{figure}
%%%%%%%%
%%%%%%%
\begin{figure}[t]
\psfrag{pop}[Bl]{$\Pi(t)$}
\psfrag{t}[c]{$t$ [units of $1/\gamma^{(0)}$]}
\includegraphics[width=8cm]{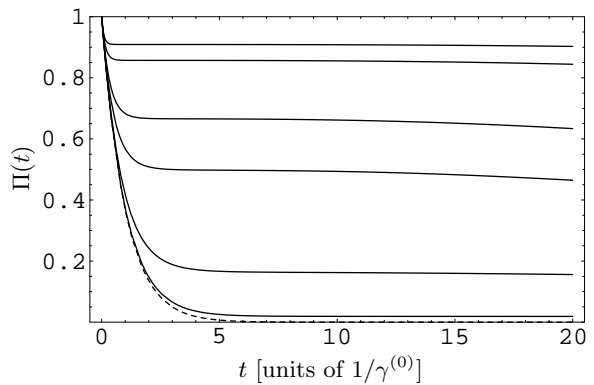}
\caption{\label{pop-gamma}Upper state population for different values of $\gamma^{(+1)} = \gamma^{(-1)}$. Starting from the top curve the chosen values are $\gamma^{(\pm 1)} = 5; 3; 1; 0.5; 0.1; 0.01$.  The dashed curve is the usual exponential decay $\exp (-\gamma^{(0)} t)$ without the low-frequency field. The other parameters are $N=1, \gamma^{(0)} = 1, \bar{\omega} = 0.1$ and $E^{(0)}(0)=1$.}
\end{figure}
%%%%%%%%
To solve the equations, the initial conditions of the atomic system have to be specified. A reasonable initial condition is $E^{(0)}(0)=1$ with all other coefficients equal to zero. This 
corresponds to the initial conditions without the low-frequency field, thus allowing a direct comparison of the system behavior with and without the low-frequency field. 

Fig. \ref{pop-omega} displays the case $N=1$ such that up to one low-frequency photon may be exchanged during an atomic transition. The dashed curve shows the usual exponential decay $(\exp(-\gamma^{(0)} t))$ one would expect without the low-frequency field as a reference. The other curves show the upper state population for different values of the low-frequency field frequency $\bar{\omega}$. The lower the frequency is, the longer is the atomic population trapped in the upper states. 
For $\bar{\omega} = 0$, the upper states are degenerate, and the upper state population is trapped permanently. As shown for the case of two upper levels in \cite{zhu} this is the expected behavior for a system with degenerate upper states. However our model system relies on the fact that photons with non-zero frequency may be exchanged during an atomic transition. Thus in experiments the frequency of the additional field may not be chosen too low. But in Fig. \ref{pop-omega} already for $\bar{\omega} = 0.1$ there is still more than 10 \% of the population in the upper states at $t=300$ (all frequencies are in units of $\gamma^{(0)}$ and all times in units of $1/\gamma^{(0)}$). 
Thus upper state population may be trapped over a long period, and the low-frequency field frequency acts as a control parameter for the trapping duration.  

In Fig. \ref{pop-gamma} the upper state population is shown for different values of $\gamma^{(\pm 1)}$. The dashed curve again is the exponential decay as a reference. The higher $\gamma^{(\pm 1)}$ is, the higher is the amount of trapped population. The effect is already visible for a very low probability of an exchange of a low-frequency photon, but increases rapidly with increasing probability for a low-frequency field-assisted interaction. Thus the relation $\gamma^{(\pm 1)} / \gamma^{(0)}$ acts as a control parameter for the amount of trapped population in the upper states.
The two parameters $\bar{\omega}$ and $\gamma^{(\pm 1)}$ can be chosen independently, as the former is the frequency of the low-frequency field while the latter depends on the intensity of the low-frequency field. Thus the decay of the upper state population can be controlled by changing the low-frequency field parameters. 

%%%%%%%
\begin{figure}[t]
\psfrag{pop}[b][c]{$|E^{(0)}(t)|^2, |E^{(\pm 1)}(t)|^2$}
\psfrag{t}[c]{$t$ [units of $1/\gamma^{(0)}$]}
\includegraphics[width=8cm]{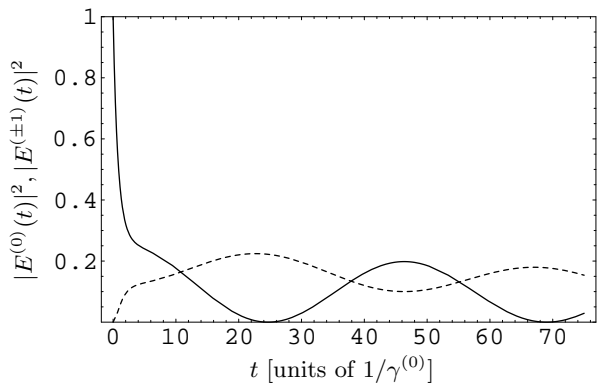}
\caption{\label{pop-seperat}Population of the different upper states. The initial conditions are $E^{(0)}(0) = 1$ and $E^{(\pm 1 )}(0) = 0$. The solid line shows the central state population, the dashed line the two other populations which are on top of each other. The parameters are $N=1, \gamma^{(0)} = 1, \gamma^{(\pm 1)}=0.5$ and $\bar{\omega} = 0.1$.}
\end{figure}
%%%%%%%%

For higher values of $N$, there is no qualitative difference to the curves shown in Figs. \ref{pop-omega} and \ref{pop-gamma}. However the more upper states there are and the higher the corresponding decay constants $\gamma^{(i)}$ are, the more effective is the population trapping.

The effects can be understood by considering the populations of the upper states separately. Fig. \ref{pop-seperat} shows the population of the three upper states $|E^{(0)}(t)|^2, |E^{(-1)}(t)|^2,$ and $|E^{(1)}(t)|^2$. At $t=0$, all the population is in the central state $\sigma_+^{(0)}\: |g\rangle |0\rangle$ (corresponding to the state $|e\rangle |0\rangle$ without the low-frequency field) due to the chosen initial conditions. Then for a short period, the center population decays exponentially as without a low-frequency field ({\it burst phase}). However, due to the quantum interference, parts of the population are transferred to the other upper states rather than remaining in the ground state. After the upper atomic levels have reached a certain superposition state, there is a distinct change in the system behavior. From now on, the total upper state population decays very slowly, while there is an oscillatory exchange of population between the different upper states ({\it quiescent phase}). If the decay constants $\gamma^{(i)}$ are chosen symmetrically, the population also distributes symmetrically.

This behavior can be explained along the lines of \cite{shapiro}. The system effectively is a superposition of overlapping resonances. Due to quantum interference, their decay may be slowed down considerably. The system also exhibits both burst phases and quiescent phases. However a crucial difference is that the system does not seem to fall into a burst phase any more after having reached a quiescent phase. Thus no external influence but the (continuous wave) low-frequency field is necessary.

It is important to note that the only condition for this scheme to work is that the low-frequency field is possibly intense and that its frequency is lower than the decay width of the upper state. Thus the scheme is not restricted to modify only a single transition. If there are more than one lower states, a single low-frequency field may suppress all transitions with decay width lower than the low-frequency field frequency simultaneously.  As in a three level $\Lambda$-system, the system then gives rise to pathway interference to each of the possible final states \cite{darkstate}. Equally, spontaneous emission may
be eliminated from a set of upper states as long as the frequency of the driving coherent field is below each of the decay rates.

{\it In summary}, we have shown that a very simple setup consisting of a single additional low-frequency field can be used to slow down the decay of an atomic two-level system considerably. The system may exchange low-frequency photons during atomic transitions. The different decay paths may be interpreted as a multiplet of upper states. Interference between the decay from these upper states is responsible for the slow down of the decay. The intensity and the frequency of the low-frequency field act as distinct control parameters for the system influencing the amount and the duration of the population trapping. A generalization of the scheme to include transitions to more than one lower atomic states is straightforward.

This work has been supported by the German science foundation (Nachwuchsgruppe within SFB 276).

\end{document}